\documentclass[aps,prl,twocolumn,superscriptaddress]{revtex4-1}
\usepackage{graphics,amsmath,graphicx,color,natbib}
\newcommand{\ave}[1]{\left\langle #1 \right\rangle}

\begin{document}
\title{Dynamic Localization of Interacting Particles in an Anharmonic Potential}

\author{M. Herrera}
\email{mherrer1@umd.edu}
\affiliation{Dept.\ of Physics and IREAP, University of Maryland, College Park, Maryland 20742, USA}
\author{T. M.  Antonsen}
\affiliation{Dept.\ of Physics and IREAP, University of Maryland, College Park, Maryland 20742, USA}
\affiliation{Dept.\ of Electrical and Computer Engineering, University of Maryland, College Park, Maryland 20742, USA}
\author{E. Ott}
\affiliation{Dept.\ of Physics and IREAP, University of Maryland, College Park, Maryland 20742, USA}
\affiliation{Dept.\ of Electrical and Computer Engineering, University of Maryland, College Park, Maryland 20742, USA}
\author{S. Fishman}
\affiliation{Physics Dept., Technion-Israel Institute of Technology, Haifa 32000, Israel}
\date{\today}

\begin{abstract}
We investigate the effect of anharmonicity and interactions on the dynamics of an initially Gaussian wavepacket in a weakly anharmonic potential.  We note that depending on the strength and sign of interactions and anharmonicity, the quantum state can be either localized or delocalized in the potential.  We formulate a classical model of this phenomenon and compare it to quantum simulations done for a self consistent potential given by the Gross-Pitaevskii Equation.

\date{today}
\begin{description}
\item[PACS numbers]
03.75.Kk, 03.65.Ge
\end{description}
\end{abstract}
\maketitle
Atoms inside anharmonic traps display a variety of phenomena including  wavepacket spreading due to dephasing, and quantum revivals \cite{Kasperkovitz1995, Robinett2004}.  Additionally, system behavior is also strongly influenced by particle-particle interactions: for example,  theoretical  \cite{Stringari1996} and experimental \cite{Mewes1996} studies have demonstrated that interactions alter the frequency of collective modes of a Bose-Einstein Condensate (BEC).  Recently, systems with both anharmonicity and interactions have been studied.  The effects of interactions and anharmonicity on the stability of stationary states \cite{Chakrabarti2009,Zezyulin2008}, collective motion \cite{Debnath2010,Perez1997,Li2006,Ott_t2003} and dynamics of coherent states \cite{Moulieras2012} of BECs have been investigated analytically and numerically.  Additionally, experimental work has also studied the dynamics of BECs in the presence of anharmonicity \cite{Jin1996,Ott2003,Bretin2004}. 

Here we consider the quantum evolution of an initially Gaussian wavepacket in a one-dimensional trap with both interactions and anharmonicity, using the description
\begin{align}
i \frac{\partial}{\partial t  } \psi & = \left (-\frac{1}{2}\frac{\partial^2}{\partial x^2}+ \frac{x^2}{2} + V_a+V_{int} \right) \psi \label{GP} 
\end{align}
where $V_a =\beta x^4/4$, and $V_{int}=u |\psi|^2$. This can be considered as the mean field approximation for the condensate wavefunction $\psi(x,t)$ of a BEC with interactions, and is known as the Gross-Pitaevskii (GP) Equation  \cite{Gross1961,Pitaevskii1961,Pethick2002}. Here $\beta$ quantifies the anharmonicity of the trap, and $u$ is a nonlinear interaction strength, which quantifies repulsive ($u>0$) or attractive ($u<0$) interactions.  We consider the regime where $V_a, V_{int}$ are smaller than the harmonic oscillator Hamiltonian, and adopt harmonic units, $x= \bar{x}/(\sqrt{\hbar/m\omega_0 }) $, $t=\omega_0\bar{t}$, where $\bar{x}$ and $\bar{t}$ are the unnormalized units,  $\omega_0$ is the frequency of the harmonic oscillator,  $m$ is the mass, and the position and momentum operators, $x$ and $p$ satisfy $[x,p]=i$, with the normalization  $\int dx |\psi|^2=1$.  We numerically solve Eq. (\ref{GP}) using a split-step Fourier method \cite{Gardiner2000}.   One possible experimental realization is the use of a Feshbach resonance \cite{Strecker2002,Cornish2006,Deissler2010,Fattori2008} in order to sample small repulsive and attractive interaction strengths.
Consider an initial, approximately Gaussian wave-packet centered at $x=x_c$, with symmetric widths in $x$ and $p$, $\Delta_x =\Delta_p=\sqrt{1/2}$.   This state may be formed by taking the confined ground state at the center of the potential and discontinuously shifting the potential to the left by an amount $x_c$.  In the case where the potential is formed by the interference of two laser beams, this shift might be accomplished by changing the relative phase of the two beams, for example as in \cite{Buchkremer2000}.

In the absence of interactions, a wavepacket that overlaps many energy eigenstates of an anharmonic trap will dephase due to the nonuniform spacing of the relevant energy eigenstates.  A classical interpretation is that in the phase space, an initial Gaussian distribution  subtends various action tori, each with a different frequency.  Thus, different parts of the distribution function rotate in the phase space at different rates, leading to spreading of the initial distribution function.   

\begin{figure}[h]
\begin{center}
\includegraphics*[height=0.43\textwidth,angle=0,clip]{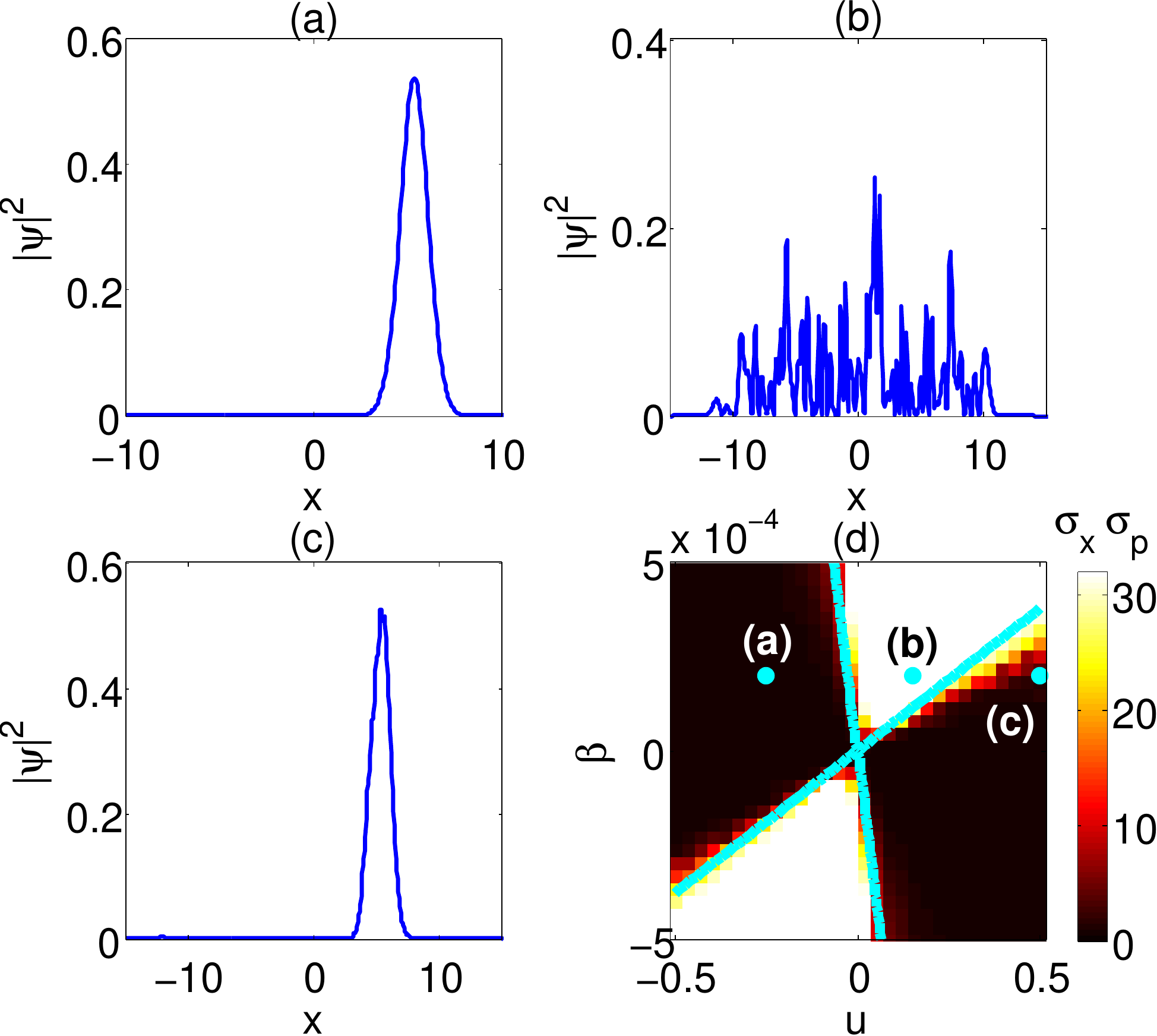}
\caption{(Color Online) (a)-(c): The position density $|\psi|^2$ at $t=1999$ for $\beta=1.89\times 10^{-4}$  with different values of $u$, (a) $u=-0.2586$,  (b) $u=0.1552$, and (c) $u=0.50$.  Note that increasing repulsive interaction strength can lead to localization.  (d): Values of $\sigma_x \sigma_p$ for different values of $u$ and $\beta$.  Bright regions indicate delocalized behavior (large $\sigma_x\sigma_p$), dark regions indicate localized behavior (small $\sigma_x\sigma_p$).  The solid lines (blue) are theoretical predictions separating regions of poor initial confinement and strong initial confinement. }
\label{heatmap_zoom_trials}
\end{center}
\end{figure}

In the presence of interactions, however, we have numerically observed what we term a ``dynamical localization" phenomenon in the time evolution of the position density of the wavepacket.  For some choices of interaction strength $u$ and anharmonicity $\beta$, the wavepacket does not spread over the trap, while for others it does.  This is illustrated in Fig. \ref{heatmap_zoom_trials} where panels (a) to (c) illustrate the position density $|\psi|^2$, obtained from solution of Eq. (\ref{GP}), at late times $(t=1999)$ for $\beta=1.89 \times 10^{-4}$ and three values of interaction strength, one attractive (a) and two repulsive (b and c).  Only in the weakly repulsive case (b) does the wave function spread throughout the trap as it would in the absence of interactions.  
Figure \ref{heatmap_zoom_trials}(d) displays the value of $\sigma_x \sigma_p$, where $\sigma_x$ ($\sigma_p$) is the standard deviation of position (momentum), averaged from $t=1899$ to $t=1999$, for various choices in $u$ and $\beta$,  and $x_c=-8$.  A localized state will have smaller values $\sigma_x \sigma_p$, shown as dark regions while a delocalized state will have larger values, shown as bright regions.  The solid (blue) straight lines in Fig. \ref{heatmap_zoom_trials}(d) are results of an approximate theory (given below) for the parameter boundaries separating localized and delocalized behavior. It is important to note that, though one might intuitively expect attractive interactions $(u<0)$ to always lead to localization, Fig. \ref{heatmap_zoom_trials} demonstrates that this is not so.  In fact, for a given value of $u$, both localized and delocalized behavior is possible, and reversing the sign of both $u$ and $\beta$ leads to the same behavior.
%
%
To gain insight into the results represented in Fig. \ref{heatmap_zoom_trials}, we introduce a classical model for the dynamical localization effect, which is motivated by our initial choice $|x_c|=8$, in terms of a classical Hamiltonian.  We note for the case of Fig. \ref{heatmap_zoom_trials}, where  the state before the shift of the potential is approximated by the ground state of the harmonic oscillator, that, after the shift of the potential, one can show that the state overlaps on the order of $|x_c|$ energy levels.  Hence, if $|x_c|$ substantially exceeds one, we expect that a classical treatment will be relevant.  Thus we start by considering the classical phase space evolution of the distribution function $f(x,v,t)$ of a one-dimensional harmonic oscillator with Hamiltonian $H_0=p^2/2 + x^2/2$. Suppose the  distribution function is initially Gaussian  (in analogy to the initial Gaussian wavepacket in the quantum case) and is displaced away from the bottom of the potential so that it is centered about a point in phase space $(x_c,0)$. In the absence of interactions and anharmonicity, this distribution coherently orbits in the phase space, since the frequency is independent of the action. We introduce as small perturbations an anharmonic potential $V_a=\beta x^4/4$, as well as a self interaction potential $V_{int}=u n(x,t)$ where $n(x,t)=\int dv f(x,v,t)$ is the density.  The Hamiltonian is then $H=H_0 + \epsilon V_a + \epsilon V_{int}$, where $\epsilon$ is an order counting parameter.  Since $V_a$ and $V_{int}$  are small perturbations to the harmonic oscillator Hamiltonian $H_0$,  the center of the phase space distribution continues to oscillate in the potential at a frequency $\Omega$ close to the unperturbed value of one.  Changing  to the action-angle variables of the unperturbed oscillator, $J$ and $\theta$, and moving to a rotating frame $\phi=\theta-\Omega t$,

the equations of motion become
\begin{align}
dJ/dt & =-\partial H/\partial \phi,  \quad d\phi/dt =\partial H/\partial J  \notag \\
H & =J(1-\Omega)+ \epsilon \beta J^2 \sin^4(\phi+\Omega t) \notag \\
& \quad + \epsilon V_{int}(\sqrt{2J} \sin(\phi+\Omega t),t), 
\label{fullmo}
\end{align}
where $p=\sqrt{2J}\cos(\theta)$ and $x=\sqrt{2J}\sin(\theta)$.  Recalling that $V_a$ and $V_{int}$ are  small perturbations to the harmonic oscillator, we note that there are two different time scales: a fast time corresponding to the center of mass motion of the distribution function, and a slow time over which the shape of the distribution function evolves.  We implement a separation of time scales $t \rightarrow  t_0,$ and $ t_1 =\epsilon t_0$, $J=J_0 +\epsilon J_1  $ and $\phi=\phi_0+\epsilon \phi_1 $.   We also expand the oscillation frequency of the center of the distribution function as $\Omega=1+\epsilon \delta\Omega $.  Averaging the resultant  equations  of motion over the period of the fast time scale, and requiring that $\ave{ \partial J_1/ \partial t_0}=\ave{ \partial \phi_1/ \partial t_0}=0$ (i.e. the correction terms do not grow secularly as functions of $t_0$) yields equations for the slow time evolution:
\begin{align}
d{J_0}/d{t_1} & = -{\partial \ave{H}}/{\partial \phi_0},  \quad {d{\phi_0}}/{d t_1} =  {\partial \ave{H}}/ {\partial J_0} \notag \\
\ave{H}&=\frac{3}{4}\beta J_0 \left (\frac{1}{2} J_0 -J_c  \right)+ \ave{V_{int}} \label{uave} 
\end{align}
where $J_c=x_c^2/2$.  The evolution of the slow-time trajectories (or the trajectories averaged over the short period) approximately follow contours of constant $\langle H \rangle$.  We have determined the value of $\delta \Omega$ by requiring that $\dot{\phi}|_{J_c}=0$ since we are in a moving frame such that the center of the distribution function is stationary, and assuming no contribution to the frequency from $\ave{V_{int}}$.  In order to average the interaction potential, we are required to average the density over an oscillation period.  Recall that the initial phase space distribution is Gaussian in $x$ and $p$, centered about some $x=x_c$, $p=0$ with width $\Delta_x =\Delta_p=\Delta=\sqrt{1/2}$.  We now evaluate $\langle V_{int} \rangle$ for the early time evolution of the phase space distribution, i.e., when that distribution is still approximately Gaussian.  Noting that the Gaussian distribution does not change appreciably over one oscillation period, allows us to average $V_{int}$, leading to
\begin{align}
\ave{V_{int}}& =\frac{ u}{2\pi\sqrt{\pi}} \int_{0}^{2\pi} dt_0 \exp[-(\sqrt{2J} \sin(\phi_0+ t_0)  \notag\\
& \quad-\sqrt{2J_c}\sin(t_0))^2] .
\label{vave}
\end{align}

\begin{figure}[h]
\begin{center}
\includegraphics*[height=0.43\textwidth,angle=0,clip]{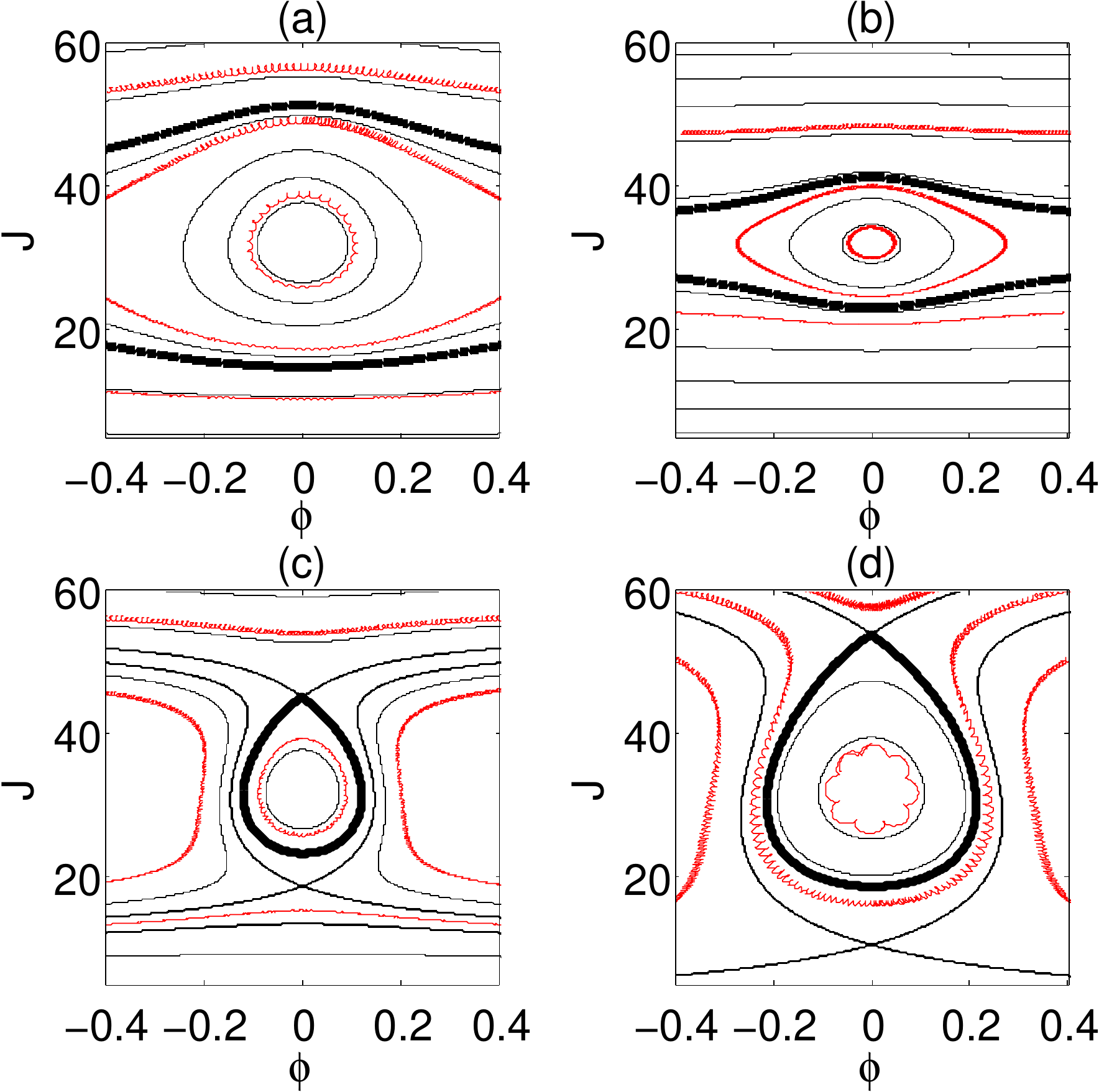}
\caption{(Color Online) Contours of $\ave{H}$ (black curves) for various values of interaction strength $u$ (a)=-0.20, (b)=-0.02, (c)=0.10, (d)=0.5 with $\beta=2 \times 10^{-4}$.  Separatrices are shown in bold, separating free streaming and confined trajectories.  In red are numerically calculated trajectories under the influence of an external potential of the form in Eq. (\ref{vave}).  The trajectories consists of short time (fast oscillations) and long time behavior.} 
\label{sos_paper}
\end{center}
\end{figure}
Numerically calculating the average of the potential, we can then estimate $\langle H \rangle$.  Figures \ref{sos_paper}(a)-(d) plot contours of $\langle H \rangle$ in black  for different choices of interaction strength $u$ and anharmonicity $\beta$, with $J_c=32$, $|x_c|=8$.

There is a difference in the nature of the phase space when $u$ and $\beta$ have opposite signs,  Figs. \ref{sos_paper}(a) and (b), and when $u$ and $\beta$ have the same sign,  Figs.  \ref{sos_paper}(c) and (d). From Eqs. (\ref{uave}) and (\ref{vave}), we see that reversing the sign of both $u$ and $\beta$ leaves phase space unchanged. For $u$ and $\beta$ of opposite signs, a fixed point is present at $(J,\phi)=(J_c,0)$, and a separatrix (shown in bold) separates free streaming trajectories from trajectories orbiting the fixed point.  In contrast, for $u$ and $\beta$ of the same sign,  while an  elliptic point is still present at $(J,\phi)=(J_c,0)$, additional hyperbolic points typically appear on the $\phi=0$ axis.  A separatrix (bold) again separates trajectories which orbit about the fixed point from those that do not, and, the trajectories immediately outside the separatrix are swept away from the region of the fixed point to values of higher $J$.   Note that the size of the region of confined trajectories increases for larger values of $|u/\beta|$.  Plotted in color in Fig. \ref{sos_paper} are numerically calculated trajectories with $ V_{int}$ given by Eq. \ref{fullmo}.  As can be seen, the trajectories have both a short time motion (fast oscillations) and long time motion in the phase space, with the long time motion given approximately by the contours of $\langle H \rangle$.

We interpret the formation of a classical elliptic region around $(J,\phi)=(J_c,0)$ as arising from a nonlinear resonance \cite{Buchleitner2002} between the integrable motion of the anharmonic well and a periodic driving term.  This  has been studied quantum mechanically, for example, in the rovibrational modes of diatomic molecules \cite{Shapiro2007} and in the dynamics of Rydberg atoms \cite{Henkel1992, Maeda2007}.  One important difference, is that in this example, the driving is not due to an external field.  Rather, the condensate/wavepacket drives \emph{itself} on resonance because $V_{int}=u n(x,t)$.  As we discuss below, this feedback between the density and $V_{int}$ can lead to dynamics which can either be localizing or delocalizing.

\begin{figure}[h]
\begin{center}
\includegraphics*[height=0.43\textwidth,angle=0,clip]{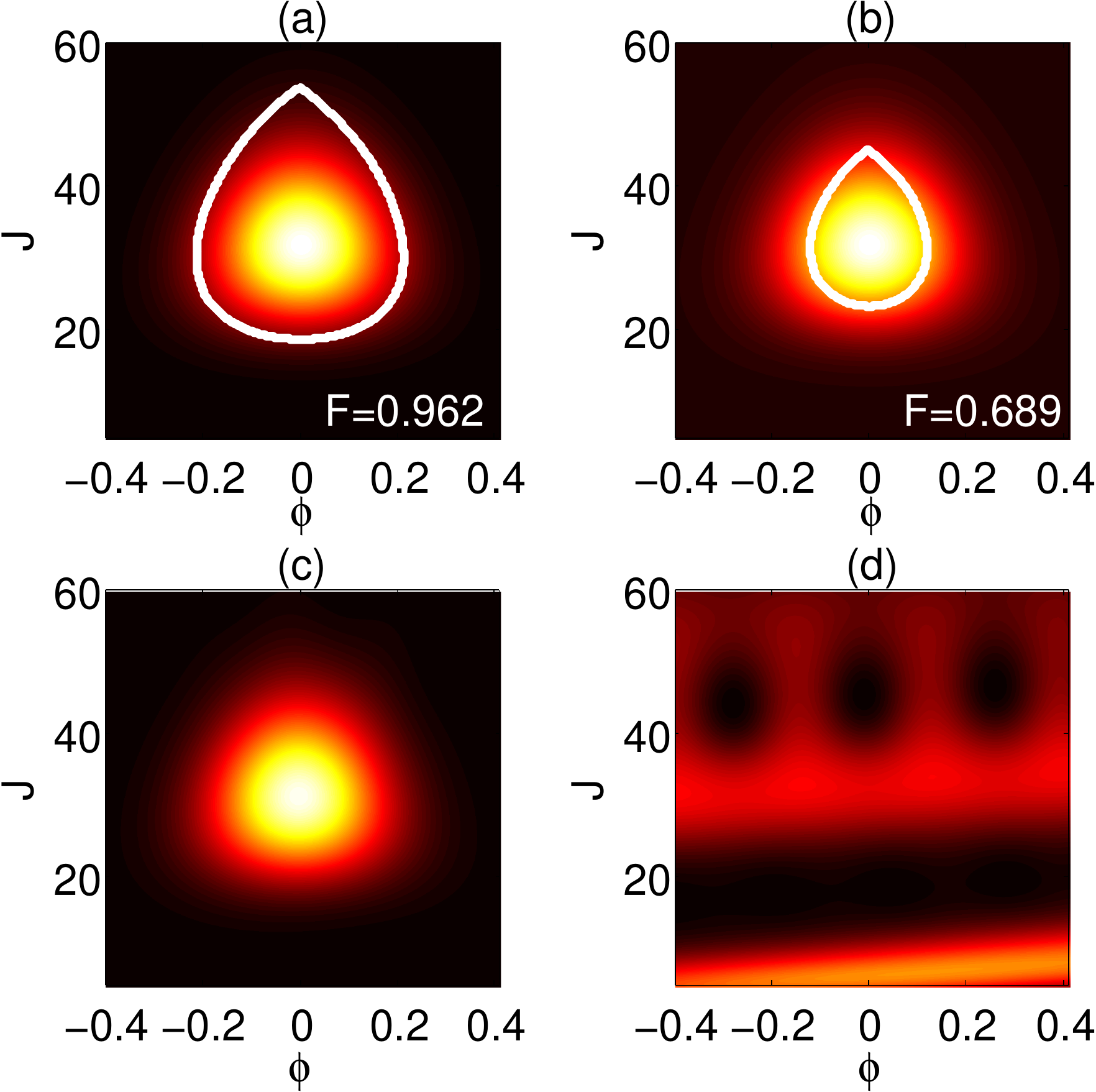}
\caption{(Color Online) The Husimi functions at $t=0$ (a)-(b) , and at $t=1999$ (c)-(d)  under the influence of the GP equation. In (a) and (b), the classical separatrix from the early time model is plotted in bold.  In  (a) $u=0.5$, $\beta=2\times 10^{-4}$, the initial condition is well contained inside the classical separatrix found with the early time model .  In  (b) $u=0.1$, $\beta=2\times 10^{-4}$, however, the initial condition extends well beyond the classical separatrix.  (c) demonstrates that the initially well contained wavepacket continues to remain localized at late times, while  (d) demonstrates that the poorly contained wavepacket has spread in the potential (color scale narrowed to show detail).  (In (c) we have applied a small shift $\Delta\phi=0.31$ to recenter the Husimi function on $\phi=0$. )} 
\label{Uave}
\end{center}
\end{figure}

Figure \ref{sos_paper} is constructed using a prescribed oscillating Gaussian potential.  We now replace it with the self consistent interaction potential $V_{int}=u n(x,t)$.  The contours of $\langle H \rangle$ found from Eqs. (\ref{uave}) and (\ref{vave}) will only describe classical trajectories for early times unless the wavefunction retains its localized character.  To this end we consider the size of the confining region compared to the extent of the Wigner function (the quantum analog to the phase space distribution, which is equal to the Gaussian distribution function at $t=0$).

Building upon insight gained from the classical model, it is reasonable to expect that if an initial Wigner function is well confined, that is,  enclosed in the classical separatrix denoting orbiting trajectories, the Wigner function will remain localized in the phase space at later times.  Similarly, if the Wigner function, is not well confined in the separatrix, then at later times it will become delocalized in the phase space. Qualitatively, as the wavepacket leaves the confining region, the strength of $V_{int}$ decreases.  This altering of the potential leads to time dependent behavior with more of the wavepacket leaving the confining region, and coexistence of a delocalized component of the state, and a localized component (most easily seen in the Husimi function representation) at lower values of action.

For example, in Figs. \ref{Uave}(a) and \ref{Uave}(b) we plot the initial Husimi function $Q(J,\phi)$ (equivalent to the Wigner function smoothed over a Gaussian function in order to assure non-negative values for all times), in the $(J$, $\phi)$ coordinates at $t=0$ for various values of $u$ with $\beta=2\times 10^{-4}$.  Plotted in bold is the classical separatrix from the Gaussian model described previously, for the figures' respective values of $u$ and $\beta$; in Fig. \ref{Uave}(a) $u=0.5$ and $\beta=2\times 10^{-4}$, while in Fig. \ref{Uave}(b) $u=0.1$ and $\beta=2\times 10^{-4}$.  We note that in Fig. \ref{Uave}(a), the Husimi function is still relatively localized inside the classical separatrix at late times,  Fig. \ref{Uave}(c).  However, in Fig. \ref{Uave}(b), where the initial Husimi extends beyond the classical separatrix, the Husimi function it not as well localized inside the classical separatrix at late times  Fig. \ref{Uave}(d). The quantity $F$ appearing in Figs. \ref{Uave}(a) and \ref{Uave}(b) is the fraction of the initial Wigner function which lies inside the confining separatrix, and reflects the difference in initial confinement between the two cases.

We test the separatrix expectation numerically by choosing different values of $u$ and $\beta$, and calculating the quantum evolution of an initially Gaussian wavepacket centered at $(x,p)=(-8,0)$.  Recall that the classical model demonstrates that the size of the confining region is dependent on the sizes and signs $u$ and $\beta$. Using our classical model, we  find for given values of $u$ and $\beta$ the fraction of the initial Wigner function that is contained in the confining separatrix, $F(u,\beta)$. Figure \ref{heatmap_zoom_trials}(d) plots in color (bold lines) contours of $F(u,\beta)$ corresponding to $F=0.9$. Note that regions yielding poor initial, classical confinement, $F<0.9$ correspond to delocalized wavepackets at late times, while regions yielding substantial initial confinement, $F>0.9$, correspond to localized wavepackets at late times.  Additionally, we have seen good agreement between localized (delocalized) numerical solutions and well (poorly) confined regions for other values of initial displacement that we have considered, $|x_c|=6$ and $9.5$.  It should be noted that this model produces similar behavior to numerical \cite{Moulieras2012} and experimental \cite{Ott2003} results reported previously for $u>0$ in the strongly interacting regime.  While keeping $u$ fixed, as the effect of anharmonicity is increased (in this case by making $\beta$ more positive, or in the previously reported cases by increasing the value of $|x_c|$), the wavepacket  becomes delocalized in the phase space, as seen in Fig. \ref{heatmap_zoom_trials}.

We note that one of the interesting features of our model is the invariance of the appearance of localization under simultaneous sign changes of $u$ and $\beta$.  This is reflected in both the numerical results of the GP Equation, Fig. \ref{heatmap_zoom_trials}(d), and the topology of classical phase space, Fig. \ref{sos_paper}.  The tendency to form a localized state is strongest when $u$ and $\beta$ have opposite signs: that is, when the trap oscillation frequency increases with action $(\beta >0)$ and interactions are attractive $(u<0)$, or oscillation frequency decreases with action $(\beta <0)$ and interactions are repulsive $(u>0)$.  This latter case is an analog of the ``negative mass instability" \cite{ [{}] [{ and references contained therein.}]  Fedele2008} that leads to bunching of a charged particle beam undergoing circular motion when the angular frequency decreases with energy.  Particles leading the bunch are repelled by the bunch, gain energy, lower their rotation rate, and fall back towards the bunch.  By this process, a beam with a uniform distribution of rotation phases will spontaneously form a bunch.  Likewise, in a trap an initial wave function that is delocalized will spontaneously tend to bunch.  

Finally, we note that the interaction values considered here are accessible in experiments.  One possible experimental realization is a lattice of one dimensional ``tubes" formed by the interference of multiple lasers \cite{Fertig2005}. In our units, the required  s-wave scattering length $a_s$ is related to the interaction parameter $u$  as $a_s=u \omega_0 \delta/(2 N \omega_{\perp})$, \cite{Gardiner2000}, where $N$ is the number of atoms per tube, $\omega_{\perp}$ is the frequency in the transverse direction, and ground state width $\delta=[\hbar/(m\omega_0)]^{1/2}$ .  Taking $\omega_{\perp}/(2\pi) \sim 49$ KHz (which is reasonable with light of wavelength $\lambda =1064$ nm and the atomic species $^{39}$K) $\omega_0/(2\pi)=2$ KHz, $N=100$, and $u=0.5$ leads to $a_s \sim 0.036$ nm.  Recently, a group \cite{Deissler2010,Fattori2008} used the Feshbach resonance in $^{39}$K to achieve $a_s$ as small as $5.29\times 10^{-3}$ nm .

\emph{Acknowledgements--}
We thank Steve Rolston for very useful discussions.  M.H. was supported by the Department of Defense (DoD) through the National Defense Science and Engineering Graduate (NDSEG) Fellowship. Further support was provided by the US-Israel Binational Science Foundation (BSF).

%

\end{document}